\documentclass[sigconf]{acmart}

\AtBeginDocument{%
  }

\copyrightyear{2023}
\acmYear{2023}
\setcopyright{acmlicensed}\acmConference[KDD '23]{Proceedings of the 29th ACM SIGKDD Conference on Knowledge Discovery and Data Mining}{August 6--10, 2023}{Long Beach, CA, USA}
\acmBooktitle{Proceedings of the 29th ACM SIGKDD Conference on Knowledge Discovery and Data Mining (KDD '23), August 6--10, 2023, Long Beach, CA, USA}
\acmPrice{15.00}
\acmDOI{10.1145/3580305.3599778}
\acmISBN{979-8-4007-0103-0/23/08}

\usepackage{algorithm}
\usepackage{algpseudocode}
\usepackage{bm}
\usepackage{hyperref}
\usepackage{siunitx}
\usepackage{balance}
\newcommand{\zx}{\boldsymbol\xi}

\begin{document}
\setlength{\abovedisplayskip}{0.2cm}
\setlength{\belowdisplayskip}{0.2cm}

\setlength{\parskip}{0cm}
\setlength{\parindent}{1em}
\title{Balancing Approach for Causal Inference at Scale}

\author{Sicheng Lin}
\authornote{Three authors contributed equally to this research.}
\affiliation{%
  \institution{Snap Inc.}
  \city{Santa Monica}
  \country{USA}
  \postcode{90405}
}
\email{slin@snap.com}

\author{Meng Xu}
\authornotemark[1]
\affiliation{%
  \institution{Snap Inc.}
  \city{Santa Monica}
  \country{USA}
  \postcode{90405}
}
\email{mxu@snap.com}

\author{Xi Zhang}
\authornotemark[1]
\affiliation{%
  \institution{Snap Inc.}
  \city{Santa Monica}
  \country{USA}
  \postcode{90405}
}
\email{xizhang@snap.com}

\author{Shih-Kang Chao}
\affiliation{%
  \institution{University of Missouri}
  \city{Columbia}
  \state{MO}
  \country{USA}}
\email{chaosh@missouri.edu}

\author{Ying-Kai Huang}
\authornote{The author contributed to the paper while working at Snap Inc.}
\affiliation{%
  \institution{Realtor.com}
  \city{Santa Clara}
  \state{CA}
  \country{USA}
}
  \email{ying-kai.huang@move.com} 

\author{Xiaolin Shi}
\affiliation{%
  \institution{Snap Inc.}
  \city{Santa Monica}
  \country{USA}
  \postcode{90405}
}
\email{xiaolin@snap.com}
\renewcommand{\shortauthors}{Sicheng Lin et al.}


\begin{abstract}
   With the modern software and online platforms to collect massive amount of data, there is an increasing demand of applying causal inference methods at large scale when randomized experimentation is not viable. Weighting methods that directly incorporate covariate balancing have recently gained popularity for estimating causal effects in observational studies. These methods reduce the manual efforts required by researchers to iterate between propensity score modeling and balance checking until a satisfied covariate balance result. 
   However, conventional solvers for determining weights lack the scalability to apply such methods on large scale datasets in companies like Snap Inc. To address the limitations and improve computational efficiency, in this paper we present scalable algorithms, DistEB and DistMS, for two balancing approaches: entropy balancing \citep{H12} and MicroSynth \citep{RSK17}. The solvers have linear time complexity and can be conveniently implemented in distributed computing frameworks such as Spark, Hive, etc. We study the properties of balancing approaches at different scales up to 1 million treated units and 487 covariates. We find that with larger sample size, both bias and variance in the causal effect estimation are significantly reduced. The results emphasize the importance of applying balancing approaches on large scale datasets. We combine the balancing approach with a synthetic control framework and deploy an end-to-end system for causal impact estimation at Snap Inc.
\end{abstract}






\maketitle

\section{Introduction}
Within the flourishing era of big data, we are dealing with an explosion of available information and high dimensional data sets. Companies like Snap Inc. conduct many quasi-experimental studies with millions of observations, aiming to demonstrate the causal relationship between an intervention and tens of critical business metrics. These causal studies boost the data-driven fashion of decision making processes and have the power to pivot future strategic business decisions. They are crucial to the company yet cannot be AB tested due to the reason that we are unable to intervene users in these cases. 

In observational studies, it is common that some of the covariates are imbalanced between treated group and control group because the treatment conditions are not randomly assigned to units. Conducting a causal analysis becomes difficult when the covariates are associated with both the treatment assignment and the outcome, as the imbalances in these covariates may introduce selection bias to the treatment effect estimation. In order to adjust for covariate imbalance, the pioneering work of propensity score \cite{Rosenbaum1983} demonstrates a key point that treatment assignment is independent of observed covariates conditional on the propensity score. Based on this property, many statistical methods has been proposed using propensity score, including propensity score matching \cite{Rosenbaum1983,Rosenbaum_Rubin_matching1985}, propensity score stratification \cite{Rosenbaum_Rubin_stratification1984}, inverse probability weighting \cite{Robins1994ipw,Hirano_imbens2001_ipw}, etc. Among these methods, weighting is the one with the most attractive statistical property. First, if propensity score is correctly modeled, the weighting estimator is consistent \cite{Lunceford2004StratificationAW}. Second, nonparametric propensity score weighting estimator is efficient \cite{Hirano_efficient_ipw}. As a comparison, matching and stratification estimators do not have these properties \cite{abadie_imbens_matching_property,Lunceford2004StratificationAW}. Furthermore, weighting methods do not require explicit outcome modeling, the weights are obtained solely from user covariates in the experimental design phase. One set of weights can be used to estimate causal effects for multiple outcomes. In this regard, the weighting methods are similar to A/B testing \cite{Rubin_2008} as they are considered more of experimental design methods than experimental analysis methods. 

Despite the solid theoretical foundation of the propensity score weighting method, in practical usage the performance of this approach is sensitive to model misspecfication. When applying propensity score weighting methods, applied researchers tend to alternately tune propensity score model and conduct balance test until a satisfied balacing result \cite{ImaKinStu08}. Entropy Balancing (hereafter EB) \cite{H12}, MicroSynth \cite{RSK17} and several other methods \cite{Zubizarreta2015, IR14, Z19} have been proposed recently to overcome the problem  by directly balancing covariates instead of modeling propensity scores. In this paper we refer to this class of methods as \textbf{balancing approach}. In general, balancing approach works better than propensity score modeling approach by achieving better balance among covariates. It is also shown that EB has doubly robust property under linear model specification \cite{ZP17}.

We rely on balancing approaches for solving causal inference problems because a set of weights can be used repeatedly for multiple business metrics and balancing approaches normally have smaller bias than propensity score modeling approaches. However we find that the current balancing approaches lack the computational efficiency to handle such big data. The number of units and dimensions in our applications can be as large as 100 millions and more than 1000 dimensions, respectively, and the overall data size can reach several tera-bytes. The majority of the existing causal inference software packages are designed for problems where the data set can be fit into one single machine. They are implemented mostly in data-science friendly languages such as Python or R, rarely to be found using distributed framework that suitable for large scale data. A Stata package \texttt{Ebalance} \cite{ebalance2013} and a R package \texttt{ebal} \cite{ebal_r_package} are the first software to find the entropy balancing weights. The \texttt{microsynth} \cite{ms_r_package} package, implemented in R, finds MicroSynth weights by solving a quadratic programming problem. Later on \texttt{empirical\_calibration} (hereafter EC) \cite{wang2019python}, a Python library developed by Google that can solve both entropy objective and quadratic objective, improves the computation speed of \texttt{ebal} by 22 times. Aside from balancing approaches, other modern causal inference software such as \texttt{EconML} \cite{econml} and \texttt{DoWhy} \cite{dowhypaper} by Microsoft, \texttt{CausalML} \cite{chen2020causalml} by Uber, \texttt{CausalImpact} \cite{causal_impact} by Google, create a rich array of causal tooling yet are limited by the memory constraint from computer machine. There is a lack of mature software dedicated for large scale causal inference tasks \cite{wong2020computational}. 

In order to scale up causal inference methods, Netflix invested in data compression strategies \cite{netflix_compression_2021} by extracting the summary statistics from original datasets. They find this strategy is able to losslessly estimate parameters from a large class of linear models \cite{netflix_compression_2021}. Several other works (e.g. \cite{logistic_spark, bluhm_cutura_2022}) attempt to use parallel computing to scale up linear models including OLS, logistic regressions etc. There is an absence of studies to scale up more advanced causal methods like balancing approaches. We design algorithms that directly utilize distributed computing frameworks to solve the optimization problems in balancing approaches. To the best of our knowledge, this is the first work that scales balancing approaches up to large datasets. 

The contribution of this paper is twofold. First, we present two distributed computing algorithms, which we name as DistEB and DistMS, to solve two well-known balancing approaches: Entropy Balancing and MicroSynth.  We exploit the dual function of the optimization problem in both approaches. We use MapReduce framework to calculate the gradient of Lagrange multiplier and apply iterative gradient based solvers. Our algorithm scales linearly with the number of observations as well as the number of covariates. It is feasible to estimate causal effects up to 100 millions units and more than 1000 dimensions. Second, we apply our methods and find that larger scale datasets significantly reduce both the bias and the variance in causal effect estimations, highlighting the importance of scalability of causal inference methods.

The rest of this paper is structured as follows. In Section \ref{sec:preliminaries} we introduce the causal inference assumptions and the optimization problems in balancing approach. In Section \ref{sec:algorithm} we describe our algorithms DistEB and DistMS in detail. Section \ref{evaluation} evaluates the performance of EB and MicroSynth as well as the scalabity of our algorithms. We discuss the applications of our algorithms at Snap Inc. in Section \ref{sec:application} and conclude in Section \ref{sec:discussion}.

\section{Preliminaries}\label{sec:preliminaries}
\subsection{Causal Inference with Observational Data: Setup and Assumptions}
Let us consider the following causal inference problem in observational studies. In a random sample of size $N$ drawn from a population, each unit $i$ receives a binary treatment assignment $T_i$. $T_i=1$ if unit $i$ is assigned in the treated group and $T_i=0$ if unit $i$ is in the control group. The number of units in the control group and the treated group are $N_0$ and $N_1$, respectively. We observe an outcome $Y_i$ and a set of covariates $X_i=[X_{i1}, X_{i2}, \cdots, X_{id}]^\top$ for each unit $i$, where $d$ is the dimension of covariates. Assuming the Stable Unit Treatment Value Assumption (SUTVA) \cite{rubin1980}, the potential outcome $\{Y_i^{(0)}, Y_i^{(1)}\}$ does not depend on the treatment assignment of other units, i.e., no network effects. Then the observed outcome can be expressed as $Y_i=T_i Y_i^{(1)} + (1-T_i) Y_i^{(0)}$. 

The treatment effect of each unit is defined as $\tau_i=Y_i^{(1)}-Y_i^{(0)}$. In this paper we focus on Population Average Treatment Effect on the Treated (PATT) estimation,
\begin{equation}
    \tau_{\textrm{PATT}}=\mathbb{E}[\tau_i|T_i=1]=\mathbb{E}[Y_i^{(1)}|T_i=1]-\mathbb{E}[Y_i^{(0)}|T_i=1].
\end{equation}
The first expectation can be simply estimated as the sample mean of outcome in the treated group. The second term describes unobserved counterfactual and needs to be estimated using data from the control group. In a randomized controlled trial, the treatment groups are randomly sampled from a population, therefore we can use the sample mean of outcome in the control group to estimate the second term because $\mathbb{E}[Y_i^{(0)}|T_i=1]=\mathbb{E}[Y_i^{(0)}|T_i=0]$. However in observational studies, where the treatment assignment is not random, selection into treatment and the outcome may both relate to confounding factors, causing the two expectations unequal. In order to correct the selection bias caused by confounders, we adopt the assumptions of strong ignorability and overlap from \cite{Rosenbaum1983}. 

\textbf{Assumption 1 (strong ignorability)}: $(Y^{(0)}, Y^{(1)}) \perp T | X $. It implies that there are no unobserved confounders.

\textbf{Assumption 2 (overlap)}: $0<P(T=1|X)<1$. It states that every unit has positive possibility of receiving each treatment condition.

\subsection{Weighting Methods: Modeling Approach VS Balancing Approach}\label{weighting_methods}
Originated from inverse probability weighting (IPW), weighting methods attempt to find weight for each unit such that the imbalance among observed covariates are adjusted. For the PATT estimation, only units in the control group need to be assigned with weights. The weighting estimator is calculated as:

\begin{equation}
    \widehat{\tau_{\textrm{PATT}}} = \sum_{T_{i}=1}\frac{1}{N_{1}}Y_{i} - \sum_{T_{i}=0}w_{i}Y_{i},
    \label{PATT_weight}  
\end{equation}
where $w_i$ is the weight for unit $i$ and the sum of weights satisfy $\sum_{T_{i}=0}w_{i}=1$, $N_{1}$ is the number of treated units. 

There are two general approaches to calculate the weights. The modeling approach estimates the weights by modeling the probabilities of receiving treatment. The balancing approach directly minimizes covariate imbalance and the dispersion of weights. 

\subsubsection{\textbf{Modeling Approach}}\label{modeling_approach}
~\\
The modeling approach is built upon the propensity score \cite{Rosenbaum1983}, defined as the probability of treatment assignment conditional on observed covariates. In observational studies, this quantity is unknown and need to be estimated by fitting a selection model on the observed data. 

There are many ways to fit a propensity score model. The most popular one is logistic regression. Propensity score model can also be based on non-linear or non-parametric methods including regularized logistic regression, random forest, boosted regression, support vector machines, neural networks, multivariate adaptive regression splines, etc.

To estimate the PATT, the weight in IPW method is

\begin{equation}
    w_{i}^{\textrm{IPW}} = \frac{\hat{e}(X_i)(1-\hat{e}(X_i))^{-1}}{\sum_{T_{i}=0}\hat{e}(X_i)(1-\hat{e}(X_i))^{-1}},
\end{equation}
where $\hat{e}(X_i)$ is the estimated probability of receiving treatment. When the propensity scores are estimated consistently, the IPW estimator is also consistent. 

One way to improve IPW estimator is by combining the propensity score model with an outcome model to achieve doubly robustness (DR). This doubly robust procedure produces consistent estimates if either of the two models are correctly specified \cite{DR_consistency}. A DR estimator can be constructed in numerous ways \cite{Kang_2007}. A modified form of regression estimation with residual bias correction employed by \cite{ZP17} is

\begin{equation}
    \widehat{\tau_{\textrm{PATT}}^{\textrm{IPW,DR}}} = \sum_{T_{i}=1}\frac{1}{N_{1}}(Y_{i}-\hat{g_0}(X_i)) - \sum_{T_{i}=0}w_{i}^{\textrm{IPW}}(Y_{i}-\hat{g_0}(X_i)),
    \label{ipw_dr}
\end{equation}
where $\hat{g_{0}}$ is an outcome model based on the data in the control group. 

There are several limitations of the modeling approach. First, when the propensity score model is correctly specified, the IPW weights only balance covariates in expectation but not in finite sample. In any particular finite sample, it could be difficult to balance covariates. Second, balance check is necessary for the modeling approach. If covariates remain imbalanced between the treatment groups after IPW adjustment, we need to modify the propensity score model to achieve satisfactory balance. This adjustment procedure involves in model selection and hyperparameter tuning. With big data, it could be very expensive to look for the model to achieve the best balance. Because of the first limitation, the model achieves the best balance in a finite sample is not necessarily the correctly specified model or the model with the smallest bias (see discussion in Section \ref{evaluation_dgp}). It means the expensive model selection and tuning process does not guarantee satisfactory performance. The doubly robust IPW estimator in Equation \ref{ipw_dr} has much better performance than the IPW estimator in our simulated data. However, when we have a large number of outcome variables, we need to model them separately. The doubly robust IPW method could be expensive in big data with a large number of outcome variables.

\subsubsection{\textbf{Balancing Approach}}
~\\
The balancing approach has been proposed to overcome the limitations of the modeling approach. This approach directly balance the covariates without explicitly modeling the probability of treatment. Entropy Balancing (EB) \cite{H12} is a groundbreaking method that minimize the Kullback-Leibler divergence between estimated weights and base weight. Usually the base weights are set to uniform weights and can be omitted from the objective function. The EB weights are calculated by solving the following optimization problem

\begin{equation}
\begin{aligned}
    \min &\sum_{T_{i}=0}w_{i}\log(w_{i})
    \\\text{s.t.} &\sum_{T_{i}=0}w_{i}c_{j}(X_i) = \frac{1}{N_1}\sum_{T_{i}=1}c_{j}(X_i) \ \forall j,
    \\&\sum_{T_{i}=0}w_{i}=1, 
    \\&w_{i}\geq 0 \ \forall i,
    \label{EB_equation}
\end{aligned}
\end{equation}
where functions $c_j$ are moment functions of the covariates. Substituting the solution of the above optimization problem into Equation \ref{PATT_weight}, the EB estimator of PATT is

\begin{equation}
    \widehat{\tau_{\textrm{PATT}}^{\textrm{EB}}} = \sum_{T_{i}=1}\frac{1}{N_{1}}Y_{i} - \sum_{T_{i}=0}w_{i}^{\textrm{EB}}Y_{i}.
    \label{EB_PATT_weight}
\end{equation}

By enforcing covariate balancing constraint, EB implicitly fits a linear regression model. The EB estimator is proved to be doubly robust and achieves the efficiency bound \cite{ZP17}. This indicates that if either the true propensity score model or outcome model are linear, the PATT is consistently estimated. Simulation results in Section \ref{evaluation} shows that EB has much better performance in balance and bias than IPW and similar performance to doubly robust IPW.

There are several extensions based on EB. Instead of the K-L divergence, the objective function of the minimization problem can also be other forms, including empirical balancing calibration weights with $D(w_{i},1)$ \cite{Chan2016},\footnote{$D(x,x_{0})$ is a distance measure for a fixed $x_{0} \in \mathbb{R}$ that is continuously differentiable in $x_{0} \in \mathbb{R}$, nonnegative and strictly convex in $x$.} the stable balancing weights which limits the possibility of extreme weights with $(w_{i}-1/r)^2$ \cite{Zubizarreta2015}, and a special case of stable balancing weights, MicroSynth with $\frac{1}{2}(w_{i}-1)^2$ \cite{RSK17}. If the constraints in the optimization problem \ref{EB_equation} hold, these extensions are also doubly robust \cite{ZP17}.

Another popular method of balancing approach is covariate balancing propensity score (CBPS) \cite{IR14}. In contrast to EB and its extensions, CBPS explicitly fits the propensity score model with balancing weights. This method is a combination of modeling approach and balancing approach. However, our simulation shows CBPS may not achieve satisfactory balance. It leaves the estimated PATT susceptible to bias. 

It is noteworthy to mention that the balancing approach does not limit to estimate PATT. Some recent studies extend this framework to estimate the Average Treatment Effect (ATE) \cite{Josey2021,Chen2022}.

More advanced methods of balancing approach emerged recently. Kernel balancing \cite{kernel_balancing} makes the estimated multivariate density of covariates approximately the
same for the treated and control groups by balancing the kernel matrix instead of original covariates . Energy balancing \cite{energy_balancing} aims to balance the distribution of covariates. However, these methods are much more computationally expensive than EB and its extensions therefore pose a practical limit of sample size and dimensions. 

\section{Balancing Approach at Scale} \label{sec:algorithm}
Our main objective in this paper is to provide a scalable solution for balancing approaches such as EB and MicroSynth. The convex optimization problem in balancing approaches can be summarised in Equation \ref{EB_equation} with different forms of loss function. The original paper of EB \cite{H12} and the implementation \cite{ebalance2013} suggest solving the optimization problem by Newton method with a line search on the Lagrange dual problem. Newton methods require calculations of the inverse of the Hessian matrix for the parameter update direction. However the inverse matrix calculation can be computationally expensive in high dimensional settings. Similar computational concerns arise for MicroSynth, where the objective function is a square loss and the optimization problem becomes a high dimensional quadratic programming with both equality and inequality constraints. We aim to provide a solution that scales linearly with number of units and utilize distributed computing framework such as Spark or Hadoop. 


\subsection{Entropy Balancing}\label{subsec:EB}
For notation simplicity, we consider the first moment function $c_j(X_i) = X_i$ in the following sections, the results can be applied for higher moments as well. In order to solve the EB optimization problem, we apply Lagrange multipliers $\zx\in\mathbb{R}^d$ to obtain the primal problem:

\begin{equation}
    \textrm{min} \bigg\{\mathcal{L}^{\textrm{EB}}= \sum_{T_i=0} w_i \log (w_i) + \zx^\top (\sum_{T_i=0} w_i X_i - \tilde X) + \xi_0 (\sum_{T_i=0} w_i - 1)\bigg\},
\end{equation}
where $\tilde X = \frac{1}{N_1}\sum_{T_{i}=1}X_i$ is the treatment mean of covariates, $\xi_0$ is the Lagrange multiplier on the normalization constraint. Solving $\partial \mathcal{L}^{\textrm{EB}}/\partial w_i=0$ yields solution of $w_i$:

\begin{equation}
    \hat w_i = w_i(\hat\zx)= \frac{\exp(-\hat\zx^\top X_i)}{\sum_{T_i=0} \exp(-\hat\zx^\top X_i)}. \label{eq:weights}
\end{equation}
Notice that this expression automatically satisfy the non-negative condition of $w_i\geq0$. Given $\hat w_i$, the Lagrange dual problem is 

\begin{equation}
    \min_{\zx\in\mathbb{R}^d}\bigg\{\mathcal{L}(\zx)=\log\bigg(\sum_{T_i=0} \exp(-\zx^\top X_i)\bigg)+\zx^\top \tilde X \bigg\}. \label{eq:dual}
\end{equation}

Considering the case where the data size is large and  cannot be fitted into one single machine, we seek for distributed methods to solve the optimization problem in Equation \ref{eq:dual}. While implementing large scale distributed L-BFGS solver remains an option for us, we utilize gradient descent solver for its simplicity and computation efficiency.  The gradient of the dual problem can be expressed as
\begin{equation}
    \frac{\partial \mathcal{L}}{\partial \zx} = \tilde X - \sum_{T_i=0}  \frac{\exp(-\zx^\top X_i)}{\sum_{T_i=0} \exp(-\zx^\top X_i)} X_i = \tilde X - \sum_{T_i=0} w_i(\zx) X_i.
\end{equation}
At each updating step of gradient descent, we first compute the weights $w_i(\zx)$ using MapReduce, the gradient can then be calculated by a second MapReduce step. In practice, momentum \cite{momentum1999} is used to accelerate the convergence rate. Once $\zx$ is solved from the dual problem, the final weights in Equation \ref{eq:weights} can be easily calculated using MapReduce. We present DistEB in Algorithm \ref{alg:DEB}.

\begin{algorithm}[ht] 
\caption{Distributed Entropy Balancing (DistEB)}
\label{alg:DEB}
\begin{algorithmic}[1]
\State Randomly initialize $\zx^{(0)} \gets \mathcal{N}(\mathbf{0}, \bm{I}_d)$, initialize learning rate $\alpha\gets0.01, \beta\gets0.9$.

\For{$k = 1,\ldots, K$} \Comment{$K$: the number of iterations}
\State Distributed compute weights $w_i(\zx^{(k-1)})$ defined in Equation \ref{eq:weights}. \Comment{MapReduce}
\State $\partial \mathcal{L} / \partial \zx^{(k-1)} = \tilde X - \sum_{T_i=0} w_i\big(\zx^{(k-1)}\big) X_i$ \Comment{MapReduce}
\State $\bm{v}^{(k)} \gets  - \alpha \partial \mathcal{L} / \partial \zx^{(k-1)} + \beta * \bm{v}^{(k-1)}$
\Comment{Momentum step}
\State $\zx^{(k)}\gets\zx^{(k-1)} + \bm{v}^{(k)}$  
\Comment{Gradient descent}
\EndFor
\State Compute final weights $w_i^{\textrm{EB}}=w_i(\zx^{(k)})$ for all control users. \Comment{MapReduce}
\end{algorithmic}
\end{algorithm}

We iterate through the whole data set twice at each updating step, the overall time complexity is $O(NdK)$, where $N$ is number of control units, $d$ is the dimensions of balancing covariates, and $K$ is the number of iterations. DistEB is implemented using PySpark. 


\subsection{MicroSynth}\label{subsec:MicroSynth}
MicroSynth differs from EB in the sense that it adopts a truncated square loss, so the solution would typically be found at the corner of the constraint set and lead to sparse weights. Besides, MicroSynth prevents outlying weights by punishing more on large weights than EB \cite{RSK17}. The optimization problem of MicroSynth is much more challenging than EB because it involves constraints that prevent negative weight. For simplicity, we add a constant into covariate vector $X_i'=[1, X_{i1}, X_{i2}, \cdots, X_{id}]^\top$, and the corresponding treatment mean $\tilde X' = \frac{1}{N_1}\sum_{T_{i}=1}X_i'$. Let $A=[X_1', X_2', \cdots, X_{N_0}']$ be a matrix of covariates for all users in the control group and $\bm{w}=[w_1, w_2, \cdots, w_{N_0}]^\top$ be the vector of weights, then the equality constraints in Equation \ref{EB_equation} can be expressed as $A\bm{w} = \tilde X'$.  We apply Lagrange multipliers $\zx\in\mathbb{R}^{d+1}$ like in EB and the primal problem is

\begin{equation}
    \textrm{min} \bigg\{\mathcal{L}^{\textrm{MicroSynth}}= \frac{1}{2}||\bm{w} - \mathbf{1}||_2^2 + \psi_+(\bm{w}) + \zx^\top (A\bm{w} - \tilde X') \bigg\}, \label{eq:ms_primal}
\end{equation}
where $\psi_+(\bm{w})=\infty$ if any $w_i<0$ and 0 otherwise. This is a high dimensional quadratic programming problem with both equality and inequality constraints. The number of parameters can be at scale of hundreds of millions, and it is infeasible to scale up traditional solvers such as interior point, projected gradient methods, active set, etc. We have found that dual ascent method is well suited for large scale distributed convex optimization problems. Another option that we have considered is Augmented Lagrangian methods like ADMM \cite{admm}, however we find that dual ascent is computationally more efficient.

The dual ascent method works in an iterative fashion. At each iteration we use gradient ascent to update the Lagrange dual variable, the optimal primal variable can then be recovered by minimizing the primal problem. The dual ascent method utilizes the fact that the Lagrange dual problem is always concave, unconstrained, and relatively low dimensional, so we are able to use gradient based methods in a distributed manner to solve the dual problem. Let $f(\bm{w})=\frac{1}{2}||\bm{w} - \mathbf{1}||_2^2 + \psi_+(\bm{w})$, the dual problem of Equation \ref{eq:ms_primal} is  

\begin{equation}
    \max_{\zx} \bigg\{\inf_{\bm{w}} \mathcal{L}^{\textrm{MicroSynth}} = -f^*(-A^\top \zx)-\tilde X'^\top \zx \bigg\}, \label{eq:ms_dual}
\end{equation}
where $f^*$ is the convex conjugate of $f$. Because $f({\bm{w}})$ is strictly convex, $f^*$ is differentiable. Using properties of conjugates, the derivative of Equation \ref{eq:ms_dual} is $A{\bm{w}}^+-\tilde X'$ where $\bm{w}^+={\mathrm{argmin}_{\bm{w}}} ~\mathcal{L}^{\textrm{MicroSynth}}(\bm{w}, \zx)$. Therefore the dual ascent method starts with an initial $\zx^{(0)}$ and updates $\bm{w}^{(k)}, \zx^{(k)}$ alternatively for $K$ iterations:
\begin{align}
    \bm{w}^{(k+1)} &= \underset{\bm{w}}{\mathrm{argmin}}~\bigg\{\frac{1}{2}\|\bm{w}-\mathbf{1}\|_2^2 + \psi_+(\bm{w}) + \zx^{(k)\top} (A\bm{w}-\tilde X')\bigg\}, \label{eq:da1}\\
    \zx^{(k+1)} &= \zx^{(k)} + \alpha^{(k)}(A\bm{w}^{(k+1)}-\tilde X').\label{eq:da2}
\end{align}
By calculus and projection, the problem in Equation \ref{eq:da1} has a close-form solution:
\begin{align}
    w_i^{(k+1)} &= \left\{
    \begin{array}{ll}
         1 - X_i^\top\zx^{(k)},& \mbox{ if }1 - X_i^\top\zx^{(k)}> 0 \\
         0,& \mbox{ otherwise.} 
    \end{array}\right.
\end{align}
Notice that this solution can be easily calculated using MapReduce because the weight $w_i$ only depends on the covariates of user $i$. In practice we add momentum in the dual gradient ascent step to speed up the convergence. The full detail of DistMS is described in Algorithm \ref{alg:DMS}.

\begin{algorithm}[ht] 
\caption{Distributed MicroSynth (DistMS)}
\label{alg:DMS}
\begin{algorithmic}[1]
\State Initialize $\zx^{(0)} \gets \mathbf{0}$, momentum $\bm{v}^{(0)} \gets \mathbf{0}$, learning rate $\alpha=0.01, \beta^{(0)}=1$.

\For{$k = 1,\ldots, K$} \Comment{$K$: the number of iterations}
\State  $w_i^{(k)} \gets \left\{
    \begin{array}{ll}
         1 - X_i^\top\zx^{(k-1)}, \mbox{ if }1 - X_i^\top\zx^{(k-1)}> 0 \\
         0, \mbox{ otherwise.} 
    \end{array}\right.$ \Comment{Distributed compute weights using Map}
\State $\bm{v}^{(k)} \gets \zx^{(k-1)} + \alpha(A\bm{w}^{(k)}-\tilde X')$
\Comment{Dual ascent step, MapReduce is used to compute $A\bm{w}^{(k)}$}
\State $\beta^{(k)} \gets \frac{1+\sqrt{1+4(\beta^{(k-1)})^2}}{2}$ \Comment{Increase the 2nd learning rate}
\State $\zx^{(k)}\gets \bm{v}^{(k)}+\Big(\frac{\beta^{(k-1)}-1}{\beta^{(k)}}\Big)(\bm{v}^{(k)}-\bm{v}^{(k-1)})$ \Comment{Momentum step}
\EndFor
\State Compute final weights $w_i^{\textrm{MicroSynth}}=w_i^{(K)}$ for all control users. \Comment{MapReduce}
\end{algorithmic}
\end{algorithm}

Similar to DistEB, DistMS loops through the whole data set twice in each iteration and the overall time complexity is $O(NdK)$.

\subsection{Comparison with other implementations}
Our implementations exploit the Lagrange dual problem like most of the existing implementations of balancing approaches. The difference lies in the way we solve the dual problem. \texttt{ebal} uses a Levenberg-Marquardt scheme to iteratively solve the problem \cite{ebal_r_package}. The drawback is that it requires calculation of the Hessian matrix, which is infeasible under high dimensional settings. EC directly uses \texttt{scipy.optimize.fsolve} to solve the dual problem \cite{wang2019python}. Under the hood, \texttt{scipy.optimize.fsolve} calculates the Jacobian matrix by a forward-difference approximation, then uses a modified Powell method to iteratively find the solution. This approach is faster than the Hessian matrix calculation. Our method is even faster in the sense that the gradient of the dual variable is directly computed from the data instead of a forward-difference step with a lot of extra function evaluations. Besides, our method is easier to implement in a distributed fashion. We compare the scalability of our algorithms with other implementations and find that our algorithms are faster than the existing in-memory implementations of EB and MicroSynth, and competitive with mathematically simpler methods such as IPW. The detailed results are described in Section \ref{subsec:runningtime}.

\subsection{Distributed vs in-memory}\label{subsec:in-mem}
Although our main focus is to design distributed algorithms for EB and MicroSynth, the algorithms can also be used serially and implemented in-memory. We implement in-memory version of both DistEB and DistMS in Python, utilizing NumPy's fast vectorized operations. When the data set is small and fits into a single-node machine, the in-memory version is very efficient because in-memory operations are much faster at different orders of magnitude than data shuffling between machines. However, machines with Terabytes of memories are not always available, in which case we need to distribute the workload on a cluster of workers. Moreover, when the data size is large, loading data into memory will be slow and take up a large portion of the total running time. We provide critical code snippets and other details for both the in-memory and Spark implementation of EB and MicroSynth in supplementary materials (Section \ref{exp_details}).


\section{Evaluation}\label{evaluation}
\subsection{Data Generating Process}
We use Monte Carlo simulations to evaluate the performance of different weighting methods on semi-synthetic data. We compare DistEB with DistMS, IPW based on logistic regression,\footnote{Other IPW procedures, such as machine-learning based approach like boosted regressions \cite{boost_reg_ipw} can also be applied to calculate the propensity score. However, with model selection and hyper-parameter tuning, these machine-learning based approach could be very expensive in big data. Therefore, we do not discuss the performance of these methods.} CBPS,\footnote{We use Meta's python package balance to implement CBPS. We focus on over-identified parametric CBPS which simultaneously fits a propensity score model and generates balancing weights. There is also a just-identified version of CBPS which only generates balancing weights without maximizing the treatment selection prediction. We also apply this method in simulation and find the results are very similar to the over-identified CBPS. We do not implement non-parametric CBPS which involves in very intensive calculation.} and the combination of these methods and a doubly robust (DR) process in Equation \ref{ipw_dr} by fitting the outcome variables with simple Ordinary Least Square regressions. 
To make the data generating process as realistic as possible, instead of using fully simulated data with normal distribution, we base our simulations on semi-synthetic data so that our simulations reflect 9-week real user data on the Snapchat platform. We define week 1 to week 8 as pretreatment period and week 9 as post-treatment period. 

We simulate data with 487 baseline covariates. It includes 456 continuous covariates which represents weekly user engagement of 57 major engagement metrics from week 1 to week 8. Many of the user engagement metrics are skewed and similar to the zero-inflated negative binomial distribution. We also have 31 binary covariates which refer to major user attributes in week 8. We use three representative user engagement metrics: active day, app open, time spent in week 9 as the outcome variables. The self-selected treatment assignment metric is users' OS types. Users are in the treated group if OS type is iOS and control group if OS type is Android. 

We assume that all 487 covariates could influence treatment assignment and outcome variables. We focus on three specifications of outcome models $\widehat{g_{j}}$ and selection model $\widehat{f}$: linear model, linear model with interactions, and random forest model. The detailed data generating steps are in supplementary materials (Algorithm \ref{alg:loop2}). In the linear specification of outcome models and selection model, we achieve the correct specification. No regularization parameter is added to the IPW method to prevent misspecification bias in this specification. In the other two specifications, we would have misspecification problems by applying DistEB, DistMS, IPW and CBPS and only including first moment. We tune regularizaion parameter to achieve the best covariate balance (see supplementary materials Section \ref{ipw_tune} for details). Hyperparameter tuning improves the performance of IPW in these two specifications. The purpose is to compare the performance of several weighting methods when both treatment selection model and outcome models are misspecified, since we never know the correct model in real-world practice.

We evaluate the performance of each method in three different sample size: 10K, 100K, and 1 Million to show the performance gain as sample size increases. We generate 100 datasets for each sample size by random sampling. 

\subsection{Meta Metrics Evaluation}
We compare different methods and sample sizes in bias, variance,  balance, and weight stability.

We measure the bias of each outcome variable $Y_{j}$ by \textit{Absolute Mean Bias}: $AMB_{j}=\frac{1}{100}|\sum_{s=1}^{100}\widehat{\delta_{js}}-\delta_{j}|$, where $\widehat{\delta_{js}}=\frac{\widehat{\tau_{js,PATT}}}{\overline{Y_{js,T_{i}=1}}-\widehat{\tau_{js,PATT}}}$. Since different outcome variables vary in magnitude, instead of PATT, we focus on the percentage difference between the treated group and the re-weighted control group. The ground truth percentage change $\delta_{j}$ is always 0 since we do not introduce any additional treatment effect.

We measure the variance by \textit{Standard Deviation}: $SD_{j}=sd(\widehat{\delta_{js}})$. It is noteworthy that another popular evaluation meta metric \textit{Root Mean Square Error (RMSE)} can be decomposed to bias and variance as: $RMSE_{j}^{2}=AMB_{j}^{2}+SD_{j}^{2}$. We do not report it separately.

We can further understand bias and variance by covariate balance. Balance means similarity between the treated group and the synthetic control group in pretreatment covariates. We use several popular balance metrics for weighting methods in literature (\cite{metrics_for_balance, IKS2007}) as follows: \textit{standardized mean difference (SMD)}, \textit{variance Ratio (VR)}, \textit{Overlapping coefficient (OVL)}, \textit{Mahalanobis Balance (MB)}, and \textit{Kolmogorov-Smirnov distance (KS)}. The detailed definition of these metrics are in supplementary materials (Section \ref{balance_weight_statbility}). We summarize the average balance across all dimensions. Among these balance metrics, SMD of original metrics ($X_j$) and MB measure the balance at mean. SMD of quadratic terms ($X_j^2$), SMD of interaction terms ($X_{j}X_{k}$), and VR measure the balance at second moments. OVL and KS measure the balance at the whole distribution.


Another diagnostic metric group is weight stability. Low stability or high variability in weights often means a few units receive very large weights. It leads to an unstable PATT estimator. Follow \cite{b_v_m}, we focus on the following meta metrics for weight stability measurement: \textit{Standard Deviation of Normalized Weights (SD)}, extremity of weights (measured by the maximum (\textit{MAX}) and 99th percentile (\textit{P99}) of weights), and \textit{Effective Sample Size (ESS)}. The detailed definition of these metrics are in supplementary materials (Section \ref{balance_weight_statbility}). A low level of ESS indicates the presence of a few units with extreme weights. Therefore, higher value of this metric indicates better weight stability. In our simulation we use $ESS$ divided by the total number of units in the control group to report the ratio of EES. This ratio can be used as a standalone diagnostic in practice \cite{b_v_m}. 

\subsection{Bias, Variance, Balance, Weight Stability in Different Data Generating Processes}\label{evaluation_dgp}
\begin{figure}[!h]
    \centering
    \vspace*{-3mm}
    \includegraphics[scale=0.52]{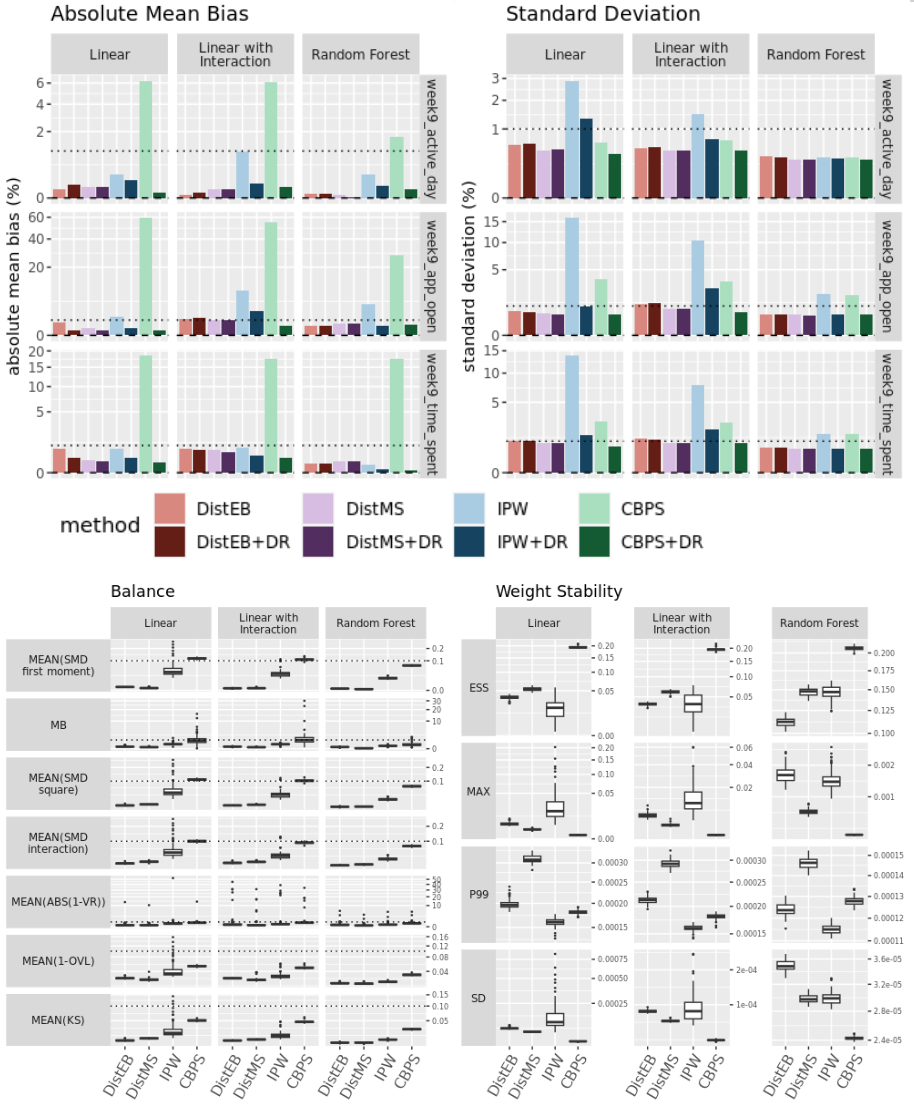}
    \vspace*{-3mm}
    \caption{Comparison of weighting methods in bias, variance, balance, and weight stability in three data generating processes (DGP). Each DGP creates 100 simulations with sample size as 100,000. DistEB and DistMS consistently outperform IPW and CBPS. The upper-left panel shows the comparison in bias measured by absolute mean bias across all simulations. The upper-right panel shows variance measured by standard deviation. The lower-left and lower-right panel show balance measured by seven balance metrics and weight stability measured by four metrics, respectively. 
    The dotted line is 1\% for absolute mean bias and standard deviation and recommended threshold in literature for balance. Y-axis is sqrt transformed for better visualization. 
    }
    \label{fig:bias_variance_dgp}
\end{figure}

We evaluate the performance of different weighting methods across three data generation scenarios, with a sample size of 100K as the primary focus. Figure \ref{fig:bias_variance_dgp} shows that no matter in correctly specified model (Linear) or incorrectly specified models (Linear with Interaction and Random Forest), DistEB and DistMS always have much smaller bias and smaller or similar variance than IPW and CBPS. When both selection and outcome models are misspecfied, DistEB and DistMS still have very small bias. It means by achieving covariate balancing, results are robust to misspecification. In real world, we always have more or less misspecification problems since we never know the ground truth data generating process. DistEB and DistMS have similar performance in bias and variance to IPW+DR and CBPS+DR. DistMS has slightly smaller variance than DistEB in all cases. DR significantly reduces the bias and variance of IPW and CBPS, but fails to enhance the performance of DistEB and DistMS. It aligns with the theory that EB and MicroSynth are implicitly doubly robust \cite{ZP17}. CBPS has worse bias but better variance than IPW in our data. It is consistent with other literature that in both correctly specified models and incorrectly specified models, compared to IPW, CBPS could have worse performance in bias but better performance in variance \cite{ZP17}.

The good performance of DistEB and DistMS can be attributed to their ability to balance covariates. It is noteworthy that we do not include higher-order moments in calculating the weights. The two balancing methods not only have good performance at mean, but also have good performance at second moment and the whole distribution. In every instance, the balance of DistEB and DistMS surpasses the recommended threshold according to literature.\footnote{We put the recommended threshold in literature in the supplementary materials. See Section \ref{balance_weight_statbility} for details.} IPW frequently falls below the threshold. Our simulations reveal that even a small imbalance can result in substantial bias. DistEB and DistMS can mitigate bias by enhancing balance, while CBPS aims to enhance balance, but fails to do so in our data.

The stability of the PATT estimator is related to weight stability. We find that even though CBPS has bad performance in bias and balance, it is the one has the best weight stability. Theoretically, EB is more likely to produce extreme weights than MicroSynth \cite{RSK17}. Our simulations confirms this statement. The performance of IPW is volatile in different simulations and is worse than DistMS and CBPS.

It's important to note that the IPW method is costly for both model selection and hyperparameter tuning, and does not guarantee the selection of a model with the lowest bias in both expectation and specific finite samples. The reason is that when the propensity score model is correctly specified, IPW only balances covariates and achieves no bias in expectations but not in any particular finite sample. Our findings show that when the data generation process is linear, tuning hyperparameters for optimal covariate balance in each simulation does not select the correctly specified model or the model with the lowest bias.

\subsection{Bias, Variance, Balance, Weight Stability by Sample Size}
\begin{figure}[!h]
    \centering
    \includegraphics[scale=0.5]{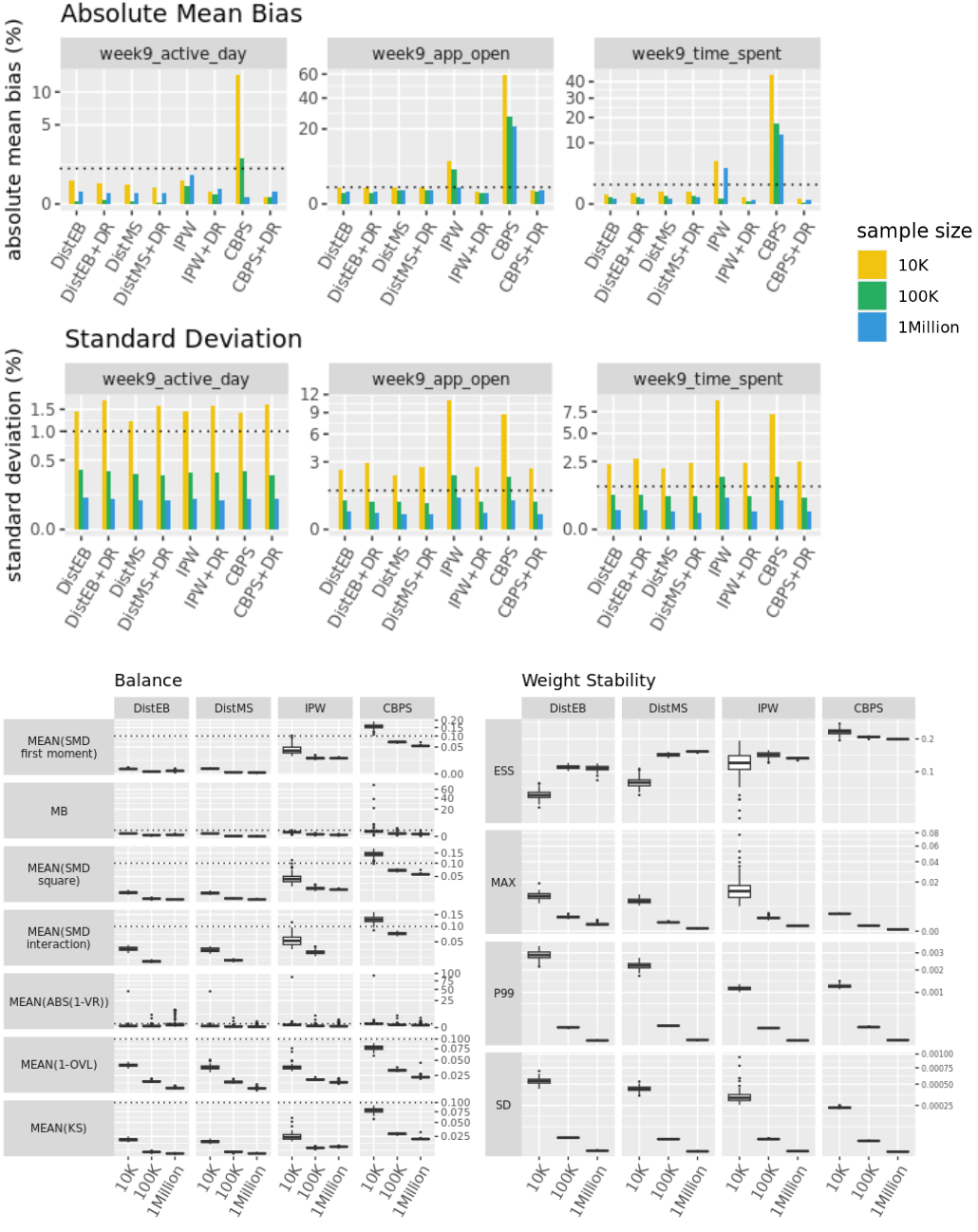}
    \caption{Comparison of weighting methods in bias, variance, balance, and weight stability for three sample sizes. With larger sample sizes, all methods performs better. The upper and the middle panel show the comparison in bias and variance, respectively. 
    The lower-left and the lower-right panel show balance measured by seven balance metrics and the weight stability measured by four metrics, respectively. 
    For each sample size, the results are derived using 100 simulations with data generating process as random forest. 
    The dotted line is 1\% for absolute mean bias and standard deviation and the recommended threshold in literature for balance. Y-axis is sqrt transformed for better visualization. 
    We do not calculate Mean (SMD interaction) for 1Million sample size because the computation is expensive.
    }
    \label{fig:bias_variance_sample_size}
\end{figure}

We evaluate the performance of various methods across three sample sizes: 10K, 100K, and 1 Million, with a focus on data generated by the random forest method. Figure \ref{fig:bias_variance_sample_size}  shows that bias and balance greatly improve from 10K to 100K but are quite similar in most cases if we compare 100K and 1 Million. Variance and some weight stability metrics always decreases as sample size increases. It means large sample size is necessary for us to get stable estimators and detect small change. 

\subsection{Running Time and Scalability} \label{subsec:runningtime}
We evaluate the in-memory and distributed version of DistEB and DistMS on Google Cloud Platform general purpose machines, details can be found in supplementary materials.

We first evaluate the in-memory version on a single-node machine and compare the performance of our method with existing open source packages including EB and MicroSynth by Google's EC package \cite{wang2019python}, CBPS by Meta's balance package, and IPW at different scales of data points (10K, 100K, 1Million). As shown in Table \ref{in_memory_rt}, compared to the existing packages of EB and MicroSynth, our in-memory version has similar or better performance in terms of running time. In general, MicroSynth outperforms EB because of faster convergence, thus requires fewer iterations. Both MicroSynth and EB significantly outperforms CBPS. IPW with L2 regularization has similar performance to MicroSynth. However, to achive the best performance, IPW requires parameter tuning, which is not included in the running time. 
\begin{table}[!h]
\caption{\label{single-node}Running Time in minutes on a Single-node Machine}
\vspace*{-4mm}
\begin{tabular}{c c c c}
\hline
\multicolumn{4}{c}{\quad\quad\quad\quad\quad\quad\quad Data Size} \\
\cline{2-4}
          &10K     & 100K  & 1Million  \\
\hline
DistEB (this work)        & 0.248       & 1.375     & 5.597 \\
EB (EC)   & 2.259  &6.066 & 124.317\\
DistMS (this work) & 0.05 & 0.142 & 0.995\\
MicroSynth (EC) & 2.342 & 0.125 & 0.99 \\
CBPS &6.411 & 44.628  & 433\\
IPW  &0.015 & 0.213 & 0.783\\
\hline
\end{tabular}

\vspace*{-5mm}
\label{in_memory_rt}
\end{table}

\begin{figure}[!h]
    \centering
    \vspace*{-10mm}
    \includegraphics[scale=0.4]{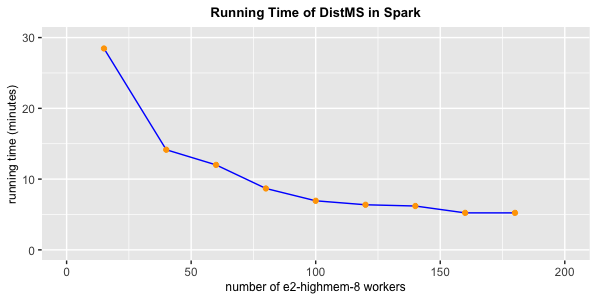}
    \vspace*{-8mm}
    \caption{Running Time of DistMS in Spark on a 10 Million dataset}
    \label{fig:running_time}
\end{figure}

For the distributed version of DistMS, we run the algorithm on a 10 Million dataset over clusters with different number of workers to demonstrate the scalability. As shown in Figure \ref{fig:running_time}, running time decreases as the number of workers increases, and stabilizes as the number of parallel tasks in Spark gets closer to the number of file partitions. Distributed computing also avoids the long data loading time on a single machine.

\section{Applications at Snap Inc.} \label{sec:application}
\begin{figure*}[!h]
    \centering
    \vspace*{-3mm}
    \includegraphics[scale=0.16]{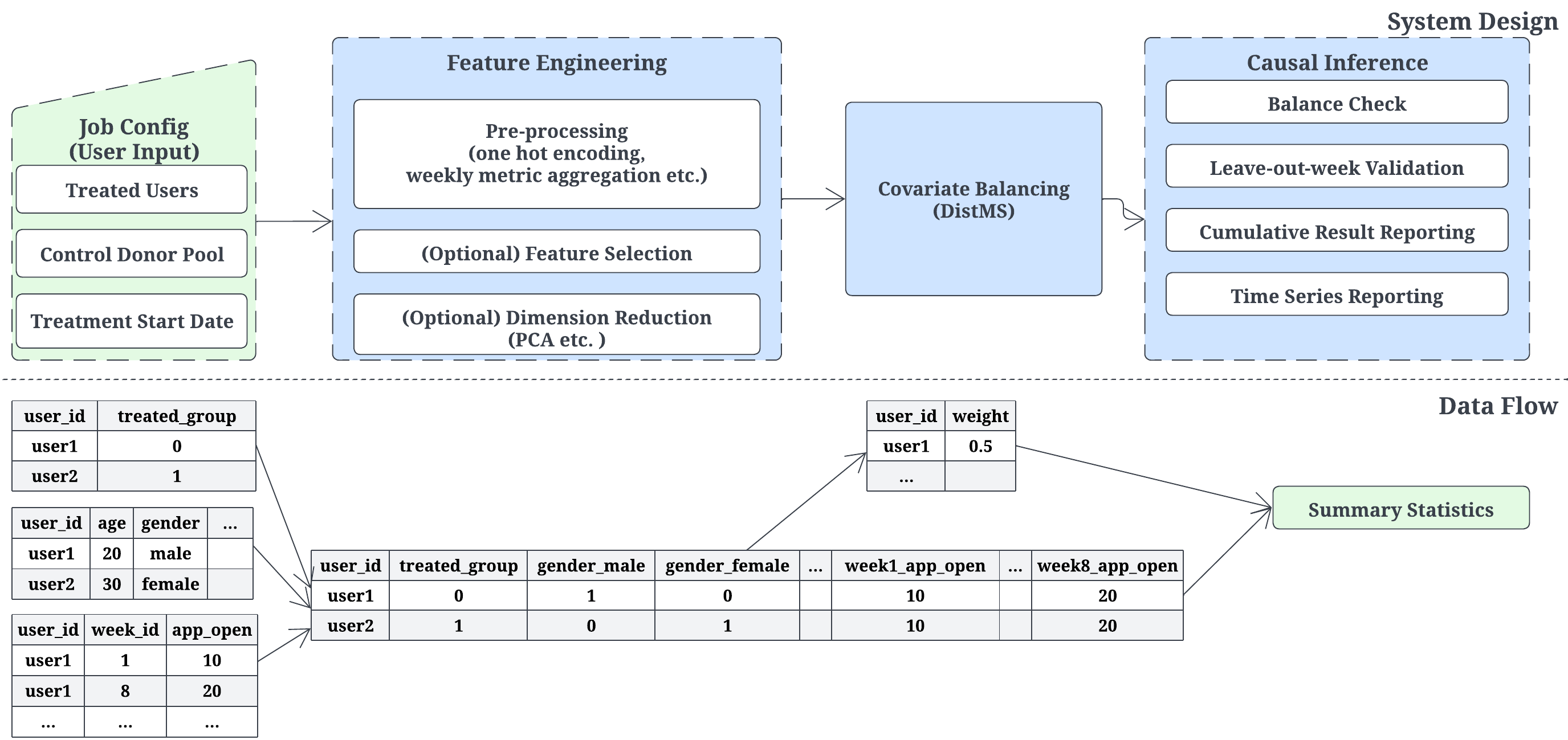}
    \vspace*{-3mm}
    \caption{End-to-end workflow for observational study analysis at Snap Inc.}
    \label{fig:system_design}
\end{figure*}

\begin{figure*}
    \centering
    \vspace*{-3mm}
    \includegraphics[scale=0.32]{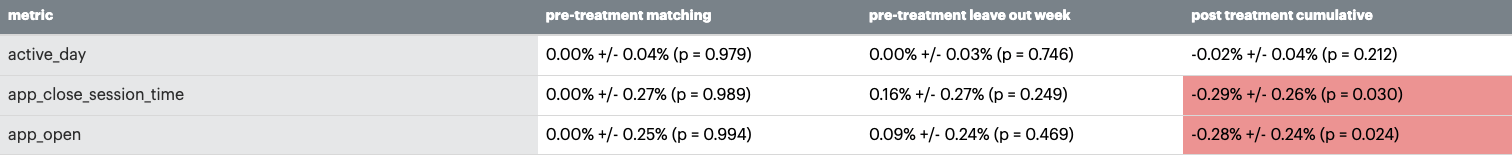}
    \vspace*{-3mm}
    \caption{Summary Report}
    \label{fig:summary_report}
\end{figure*}

\begin{figure*}
    \centering
    \vspace*{-3mm}
    \includegraphics[scale=0.29]{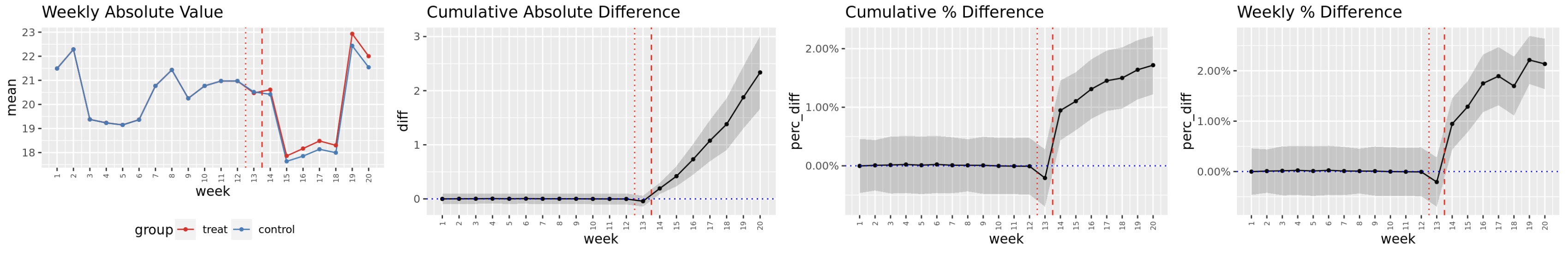}
    \vspace*{-3mm}
    \caption{Time Series Report: We have 12 pretreatment weeks, 1 validation week, and 7 post-treatment weeks in this example. The red dotted line separates pretreatment period and validation period. The red dashed line separates validation period and post-treatment period.}
    \label{fig:time_series_report}
\end{figure*}

At Snap Inc., We combine the balancing approaches and the synthetic control framework with multiple treated units \cite{A21} and design an end-to-end workflow for observational study analysis, as shown in Figure \ref{fig:system_design}. With the synthetic control framework, we balance the pretreatment time-series of multiple engagement metrics and user attributes to mitigate the bias caused by unmeasured confounders \cite{ADH2010}. We construct a synthetic control unit with millions of users in the control donor pool. The synthetic control unit serves as the counterfactual of the treated units. Considering the performance in variance, weight stability, and running time, DistMS outperforms DistEB. We productionize DistMS in the workflow. Our tool takes the user identifiers in the treated group, control donor pool and the treatment start date as inputs, then uses DistMS to generate a set of weights for users in the control donor pool. The re-weighted control group is directly comparable with the treated group. In the end, we provide a summary report containing the cumulative differences and time series. Following the best practices of the synthetic control method \cite{A21}, we run balance check on all balancing variables including pretreatment user engagement metrics and user attributes such as country, age, gender etc, which is analogical to the concept of sample ratio mismatch \cite{fabijan_gupchup_gupta_omhover_qin_vermeer_dmitriev_2019} from A/B testing. We leave one week out of matching period as the validation period, which mimics the AA period in A/B testing. Confidence intervals are calculated using bootstrap method. Users in familiarity with traditional A/B testing platform can easily interpret the results of observational studies, given the similarity between the reporting interfaces, as demonstrated in Figure \ref{fig:summary_report} and Figure \ref{fig:time_series_report}. 

This end-to-end workflow has been applied to many cases at Snap Inc. Engineers can use the tool to analyze the impact on user engagement of some events or incidents that happen naturally, such as network errors that only happen in a certain area. Data scientists can also design quasi-experiments, such as launching a marketing campaign for Snapchat in a college and measuring the impact with synthetic control. Moreover, in randomized experiments, app performance metrics such as app open latency could suffer from self-selection biases, because the metric is only reported when users open the app, as pointed out in \cite{Xie2020HowTM}. Balancing approaches can be directly applied to mitigate the biases by adjusting for relevant covariates. Other companies also face these scenarios and can apply this system to their analysis.

\section{Discussion} \label{sec:discussion}

The validity of the end-to-end system relies on certain assumptions. 

Firstly, similar to most causal inference methods with observational data, we assume no unmeasured confounders and overlap. With larger datasets, we can balance numerous confounders, mitigating biases caused by unmeasured confounders. 
In practice, we focus on pre-treatment time-series of key engagement metrics and user attributes, balancing only the first moment by default. In some cases, balancing more pre-treatment covariates can help reduce bias caused by unobserved confounders. The validation period before treatment allows us to check for strong evidence of unobserved confounders.

Secondly, we have the assumption of synthetic control framework. This assumption asserts that a good pre-treatment fit acts as a proxy for a good accounting of all observed and unobserved factors. 
Matching pre-intervention outcome time series helps control for heterogeneous responses to multiple unobserved factors. With this assumption, only unobserved confounders in the post-treatment period, which cannot be accurately predicted by pre-treatment information, can bias our results. 

Lastly, our system relies on the data generating assumption underlying the balancing approach. This assumption requires a linear true data generating process for either the selection model or the outcome model. We find that this assumption can be loosen in practice. Our simulations demonstrate that even when the true data generating process is nonlinear, our system still provides estimates with smaller bias compared to other weighting methods. 
Nevertheless, including higher moments may be helpful in reducing misspecification bias in certain cases \cite{high_order_necessary}. Incorporating a large number of balancing covariates and moments poses two challenges. Firstly, finding a solution that balances all covariates and moments becomes more difficult as more constraints are added to the optimization procedure. Secondly, including many balancing constraints increases computational costs. Therefore, applying feature selection or dimension reduction methods before the balance procedure can be helpful, and this aspect remains open for future improvement.

In summary, our findings demonstrate that the balancing approach, which directly balances confounders, outperforms the modeling approach by producing a causal effect estimator with low bias and variance. This approach is also more efficient, as it does not require modeling multiple outcome variables and adjusting hyperparameters, resulting in a doubly robust solution with low computational cost. A large sample size also reduces estimation variance. To further improve the solution, we propose scalable solvers that have faster in-memory computation than existing solvers and can be easily integrated with distributed computing frameworks. Our application combines the balancing approach with a synthetic control framework to address unmeasured confounders. This end-to-end workflow is successfully implemented at Snap Inc. to estimate causal effects using large-scale observational data.



\clearpage
\bibliographystyle{ACM-Reference-Format}
\pagebreak
\balance
\bibliography{Reference}

\appendix
\section{Details on Data Simulation}
\subsection{Steps to Generate Semi-Synthetic Data}

\begin{algorithm}[ht] 
\caption{Simulate Semi-Synthetic Data}
\label{alg:loop2}
\begin{algorithmic}[1]
\State Randomly select 2N ($N=10Million$) users from the whole population and split them equally into a training and testing dataset.
\State With the training dataset, train a regression model $\widehat{g_{j}}(X_{i})$ to predict the outcome metric $Y_{ij}$, a classification model $\widehat{f}(X_{i})$ to predict the treatment assignment. 
\State Simulate outcome metrics: generate predicted outcome $\widehat{Y_{ij}}=\widehat{g_{j}}(X_{i})+\hat{\epsilon_{ij}}$, where $\hat{\epsilon_{ij}}$ is sampled with replacement from the fitted errors of the first $N$ used for training.
\State Simulate treatment assignment: Generate propensities $Prob(T_{i}=1)=\widehat{f}(X_{i})$ for each user. Assign each of the last $N$ user into treatment or control by their predicted treatment status, which is determined by a $Bernoulli(p)$ with $p = Prob(T_{i}=1)=\widehat{f}(X_{i})$.
\end{algorithmic}
\end{algorithm}

\subsection{Specification Details}
\subsubsection{Specification 1. Linear Model}
\begin{itemize}
    \item Outcome models are fitted with linear regression, with 487 covariates.
    \item Treatment selection model is fitted with binary logistic regression, with 487 covariates.
\end{itemize}
\subsubsection{Specification 2. Linear Model with Interactions}
We interact one user attribute, which is user persona, with all other covariates. The new feature set contains 1458 covariates. 
\begin{itemize}
    \item Outcome models are fitted with linear regression, with 1458 covariates.
    \item Treatment selection model is fitted with binary logistic regression, with 1458 covariates.
\end{itemize}
\subsubsection{Specification 3. Random Forest}
\begin{itemize}
    \item Outcome models are fitted with RandomForestRegressor on 487 covariates. Number of Trees is 10. Max Depth is 10.
    \item Treatment selection model is fitted with RandomForestClassifier on 487 covariates. Number of Trees is 10. Max Depth is 15. 
\end{itemize}

\section{Details of Balance and Weight Stability Metrics}\label{balance_weight_statbility}
The \textit{standardized mean difference (SMD)}: $SMD_{d}=\frac{|\Delta_{d}|}{\sqrt{(s_{d1}^2+s_{d0}^{w2})/2}}$, where $\Delta_{d}$ is the difference in means of covariate $d$ between treated group and the re-weighted control group, $s_{d1}^2$ is the sample variance of $X_d$ in the treated group, $s_{d0}^{w2}$ is the weighted sample variance of $X_d$ in the re-weighted control group. To summarize the performance of all dimensions, we further calculate $\overline{SMD_{first\:moment}}$, $\overline{SMD_{square}}$, and $\overline{SMD_{interaction}}$ to represent the mean $SMD$ of all covariates, all quadratic terms of the continuous variables, and all possible interaction terms. Even though we do not add quadratic and interaction terms in the weighting process, we are interested in the balance performance of different methods in these terms. Most researchers consider balance to be achieved when SMD is below 0.1. It is an arbitrary threshold.

The \textit{Mahalanobis balance (MB)}: $MB=(\overline{X_{1}}-\overline{X_{0}^{w}})^{'}\sum^{-1}(\overline{X_{1}}-\overline{X_{0}^{w}})$, where $\overline{X_{1}}$ is the vector of covariate means in the treated group, $\overline{X_{0}^{w}})$ is the vector of weighted covariate means in the re-weighted control group, and $\sum^{-1}$ is the sample variance-covariance matrix of covariates. A Mahalanobis Distance of 1 or lower shows that the point is right among the benchmark points.\footnote{https://www.theinformationlab.co.uk/2017/05/26/mahalanobis-distance/}

The \textit{variance Ratio(VR)}: $VR_{d}=\frac{s_{d0}^{w2}}{s_{d1}^{2}}$. Closer values to 1 indicate better balance in the second central moment. To make it comparable with other balance metrics, we generally consider $|1-VR_{d}|$. To summarize the performance of all dimensions, we calculate $\overline{|1-VR_{d}|}$ which is the mean of $|1-VR_{d}|$ across all dimensions. For good balance, the variance ratio should be between 0.5 and 2 \cite{rubin2001,Stuart2010}. In our simulations, we use 0.5 for $\overline{|1-VR_{d}|}$. 

The \textit{Overlapping coefficient (OVL)} is the proportion of overlap in the covariate distributions between two groups. It is calculated by finding the area under the minimum of both curves.

The OVL value is always between 0 and 1. Closer to 1 means better balance. To make it comparable with other metrics, we focus on $1-OVL$. We have $1-OVL_{d}$ for each dimension. To summarize the performance of all dimensions, we calculate $\overline{1-OVL_{d}}$ which is the mean of $1-OVL_{d}$ across all dimensions. There is no recommended threshold in literature. We use 0.1 for $\overline{1-OVL_{d}}$. 

The \textit{Kolmogorov-Smirnov distance (KS)} is the maximum vertical distance between two cumulative distributions. Some literature use 0.1 as the balance threshold for KS \cite{KS_threshold}.

The \textit{Standard Deviation of Normalized Weights (SD)}: $SD=sd(w_{i0})$ with $\sum(w_{i0})=1$. It is the moment-based measures of variability. Lower value indicates better weight stability.

The extremity of weights: measured by the maximum (\textit{MAX}) and 99th percentile (\textit{P99}) of weights. Lower value indicates better weight stability.

The \textit{Effective Sample Size (ESS)}: $ESS=\frac{(\sum w_{i0})^2}{\sum w_{i0}^2}$. Higher value indicates better weight stability.

\section{Further Details on Experiments}
\label{exp_details}
We provide critical code snippets for both the in-memory and Spark implementation of EB and MicroSynth here:
\url{https://gist.github.com/xizzzz/02caadbaf760510e267cc1fa9d3a7971}
\subsubsection{Learning Rate}
We initialize learning rate with 0.01, and add decay if oscillation is detected.
\subsubsection{Stopping Condition}
Heuristically we choose a threshold that makes our model achieve desirable performance as the stopping condition. We used the difference between the sum of the weights and the target below \num{1e-4}. 
\subsubsection{Training Machine Types}
We use Google Cloud Platform general purpose machines for the experiments.\footnote{https://cloud.google.com/compute/docs/general-purpose-machines}  Different machine types are chosen to fit the computation needs of each data scale. Image version is 2.0.29-debian10. We list the machine types in the following table. Data scale represents the number of rows in each data set. \\
\begin{center}
\begin{tabular}{||c c c||} 
 \hline
 Data Scale & Machine Type & Memory (GB) \\ [0.5ex] 
 \hline\hline
 10K & n1-standard-8 & 30 \\ 
 \hline
 100K & n1-standard-16 & 60 \\
 \hline
 1M & n2-highmem-32 & 256 \\
 \hline
 10M & e2-highmem-8 & 64 * number of workers \\ [1ex] 
 \hline
\end{tabular}
\end{center}

\section{IPW hyperparameter tuning}\label{ipw_tune}
In Section \ref{evaluation} we use grid search to tune hyperparameters of logistic regression in the IPW method. We tune 3 hyperparameters: number of folds in cross-fitting $k\in \{0, 3, 5\}$, regularization parameter $C \in \{0.0001, 0.0003, 0.001, 0.003, 0.01, 0.03\}$, and l1\_ratio $l \in \{0, 0.25, 0.5, 0.75, 1\}$. For each simulation setting, the best hyperparameter is chosen with the lowest standardized mean difference.

\end{document}